\documentclass[pre,epsfig,aps,superscriptaddress,showpacs,twocolumn]{revtex4}
\usepackage{epsfig}
\usepackage{latexsym}
\usepackage{graphicx}

\usepackage{amssymb,amsfonts,amsmath}

\begin{document}

\title{Structuring and sampling complex conformation space: Weighted ensemble dynamics simulations}

\author{Linchen Gong}
\affiliation{Asia Pacific Center for Theoretical Physics and Department of Physics, Pohang University of Science and Technology,
Pohang, Gyeongbuk 790-784, Korea}

\affiliation{Center for Advanced Study, Tsinghua University, Beijing 100084, China}

\author{Xin Zhou
\footnote{Author to whom correspondence should be addressed; Electronic Mail: xzhou@apctp.org} }

\affiliation{Asia Pacific Center for Theoretical Physics and Department of Physics, Pohang University of Science and Technology,
Pohang, Gyeongbuk 790-784, Korea}

\date{\today}

\begin{abstract}
Based on multiple simulation trajectories, which started from dispersively selected initial conformations, the weighted ensemble dynamics method is
designed to robustly and systematically explore the hierarchical structure of complex conformational space through the spectral analysis of the
variance-covariance matrix of trajectory-mapped vectors. Non-degenerate ground state of the matrix directly predicts the ergodicity of
simulation data. The ground state could be adopted as statistical weights of trajectories to correctly reconstruct the equilibrium properties,
even though each trajectory only explores part of the conformational space. Otherwise, the degree of degeneracy simply gives the number of
meta-stable states of the system under the time scale of individual trajectory. Manipulation on the eigenvectors leads to the classification of
trajectories into non-transition ones within the states and transition ones between them. The transition states may also be predicted without a
priori knowledge of the system. We demonstrate the application of the general method both to the system with a one-dimensional glassy potential and with 
the one of alanine dipeptide in explicit solvent.
\end{abstract}


\maketitle

\section{Introduction}
To explore the conformational space of physical systems, standard molecular-dynamics (MD) and Monte Carlo (MC) simulation methods have been
applied for decades. All the dynamic and thermodynamic information of a model system could be extracted from a long enough, ergodic simulation
trajectory. For complex systems with a rugged free-energy surface, due to the existence of lots of free energy barriers, the standard methods are
no longer promising for enough sampling, which - as a result - impulses the development of various advanced simulation
techniques~\cite{BergN1991,WangL2001b,SugitaO2000,Voter1997b,ZhouJKZR2006,BolhuisCDG2002,Elber2002,ERV2005}. Despite the success already
achieved, it is still difficult to thoroughly investigate a practically interesting system by a single trajectory simulation with these advanced
algorithms. Besides, to ensure the convergence of simulation, a few trajectories may need to be generated separately, and then the difference
between these trajectories could be measured~\cite{ThirumalaiMK1989} to sentence the ergodicity of simulation under the simulated time scale.
Thus, it is reasonable to ask whether we could benefit from the state-of-the-art distributed computation technique to explore the conformational
space with multiple parallel generated simulation trajectories. Actually, one solution, {\it e.g.}, the ensemble dynamics 
(ED) method~\cite{Voter1998,ShirtsP2001}, has already been well established to investigate the transition dynamics between specific states with
thousands of short simulation trajectories~\cite{HuangBGP2008}. With a similar technique, but different spirit, we develop a method for
systematically discovering the meta-stable states in conformational space and the transition events between them. It could also be applied to
more efficiently sample the complex conformational space.

The method is bases on three considerations. (1) Compared to the single simulation trajectory, parallel generated trajectories, which started from
dispersive initial conformations, are possible to cover a much larger area in conformational space with the same total simulation length. (2)
Ergodicity broken simulation also provides meaningful information of conformational space. Existing methods for constructing the hierarchical
state structure usually demand thoroughly a sampling of the system either energetically~\cite{StillingerW1982} or
kinetically~\cite{BeckerK1997,Wales2001,KrivovK2004,RaoC2004,GfellerRCR2007}. As a matter of fact, the state structure is an effect of
simulation time scale. With long enough observation time, the whole conformational space seems like one state. Otherwise, the sub-states
structure of the system will emerge. Thus, even though each trajectory in ED is not ergodic, their mutual relation will reflect the state
structure of the system under the time scale of the individual simulation trajectory. (3) The short ED trajectories could be combined to give out
equilibrium properties of the system. Weighted histogram analysis method~\cite{ChoderaSPSD2007} is one existing technique to combine
several trajectories by estimating the density of states in energy space. In ED, since every trajectory is generated by the same MD (or MC)
simulation, we could simply specify a statistical weight to each trajectory to reconstruct the equilibrium distribution. The weighting scheme
leads to the name of our method, weighted ensemble dynamics (WED).

Given multiple MD trajectories generated by the same algorithm and condition, WED provides a way to systematically understand the information in
existing data without requiring much foreknowledge. It classifies the trajectories into non-transition ones within the meta-stable
conformational regions (states), and transition ones between these states. The trajectories inside the same state could be combined to mimic the
intrastate equilibrium distribution, and some of the transition trajectories could be shortened to locate the well-defined transition state
ensemble. Based on the existing data, further exploration of the system could be efficiently guided by WED, and the temporally hierarchical
structure of the conformational space could be distilled out step by step.

The paper is organized as follows. The theory of WED will be introduced in Sec.~\ref{sec:theory}. The results and discussions are presented in
Sec.~\ref{sec:results}. A short conclusion is given in Sec.~\ref{sec:conclude}. Finally, the detailed simulation methods are listed in the Appendix.

\section{Theory}
\label{sec:theory} Let us consider a set of $\tau-$length, normally simulated MD (or MC) trajectories. The distribution of their initial
conformations $P_{init}(\vec{q})$ will evolve along the trajectories to the equilibrium distribution $P_{eq}(\vec{q})$ with long enough
$\tau$. Here $\vec{q}$ generally denotes the conformation coordinates of the system. The time for approaching equilibrium is dependent on the
deviation of $P_{init}(\vec{q})$ from $P_{eq}(\vec{q})$. Instead of waiting for the final equilibration, we may specify each trajectory a
statistical weight. Then, the equilibrium ensemble average of any physical quantity $A(\vec{q})$ could be estimated by
\begin{equation}
\label{eq:weight} \langle{} A(\vec{q}) \rangle_{eq}=\frac{\sum_{i} w_{i} \langle{} A(\vec{q}) \rangle_{i}}{\sum_{i} w_{i}}, 
\end{equation}
where, $w_{i}$ is the weight of the $i$th trajectory. $\langle{}\ldots\rangle_{eq}$ and $\langle{}\ldots\rangle_i$ denote the average over
the equilibrium distribution and the $i$th trajectory, respectively.

Considering the deviation of $P_{init}(\vec{q})$ from $P_{eq}(\vec{q})$, we could choose
\begin{equation}
\label{eq:originw}
\Omega_{init,eq}(\vec{q}_{i0}) =
\frac{P_{eq}(\vec{q}_{i0})}{P_{init}(\vec{q}_{i0})}
\end{equation}
as $w_{i}$. Here $\vec{q}_{i0}$ is the initial conformation of the $i$th  trajectory. For example, if the initial conformations were chosen
from the simulation with modified potential $U(\vec{q})$, we could set $w_i \propto \exp[\beta{}U(\vec{q}_{i0})-\beta{}V(\vec{q}_{i0})]$ to
reproduce the equilibrium properties of the original system with the potential $V(\vec{q})$. However, this normal weighting scheme demands the
knowledge of $P_{init}(\vec{q})$, and usually suffers from the enormous fluctuation of $\{\Omega_{init,eq}(\vec{q}_{i0})\}$ in systems with
large degrees of freedom. As a result, only a few trajectories with dominating $w_{i}$ could significantly contribute to the weighted average.

Actually, while the length of simulation trajectories, $\tau$, is not very short, the conformational space could be divided into a few
meta-stable regions, wherein a single trajectory could reach local equilibrium within $\tau$. Thus, for trajectories that started from the same
meta-stable region, the specified weights - although dependent on the initial conformations - should be approximately identical when reproducing
the equilibrium properties. Therefore, we could write
\begin{eqnarray}
w_{i} \propto \frac{\int_{\alpha(i)} P_{eq}(\vec{q}) d \vec{q}}{\int_{\alpha(i)} P_{init}(\vec{q}) d \vec{q}},
\label{eq:state-weight}
\end{eqnarray}
where $\alpha(i)$ denotes the meta-stable region, in which the $i$th  trajectory is started. Instead of explicitly using
Eq.~(\ref{eq:state-weight}), we simply construct a new ensemble of conformations $-X-$ in practice. $X$ is constituted by the conformations in
the initial $\tilde{t}$ time length of all the simulation trajectories. Within short enough $\tilde{t}$ (compared to $\tau$), each trajectory is
supposed to be still exploring the same meta-stable conformational region, leading to the quite plausible conclusion that $\int_{\alpha}
P_{X}(\vec{q}) d\vec{q} \approx \int_{\alpha} P_{init}(\vec{q}) d \vec{q}$ for all the meta-stable regions. Here, $\alpha$ denotes the
meta-stable region in consideration; $P_{X}(\vec{q})$ denotes the distribution function of the conformations in $X$ (all the distribution
functions are supposed to be normalized). Consequently, $P_{X}(\vec{q})$, instead of $P_{init}(\vec{q})$, could be used to estimate the
trajectory weights as follows,
\begin{eqnarray}
w_{i} = \langle \Omega_{X,eq}(\vec{q}) \rangle_{i^{+}} = \left\langle \frac{P_{eq}(\vec{q})}{P_{X}(\vec{q})} \right\rangle_{i^{+}}.
\label{eq:realw}
\end{eqnarray}
Here, $\langle \cdots \rangle_{i^{+}}$ denotes the average over the initial ${\tilde t}-$ length segment of the $i$th  trajectory, {\it i.e.}, 
\begin{eqnarray}
\langle A(\vec{q}) \rangle_{i^{+}} \equiv \frac{1}{\tilde t} \int_{0}^{\tilde t} A(\vec{q}_{i}(t^{\prime})) d t^{\prime},
\end{eqnarray}
for any conformational function $A(\vec{q})$. Here $\vec{q}_{i}(t^{\prime})$ denotes the conformation at $t^{\prime}$ time in the $i$th 
trajectory. Heuristically, by averaging $\Omega_{X,eq}(\vec{q})$ over the initial short segment of each trajectory, we are calculating
$\{w_{i}\}$ according to the initial regions of the trajectories, rather than solely by the initial conformations.

Although the analytical expression of $P_{X}(\vec{q})$ is unknown, considering the general relation
\begin{eqnarray}
\langle \Omega_{X,eq}(\vec{q}) A(\vec{q}) \rangle_{X} = \langle A(\vec{q}) \rangle_{eq}
\end{eqnarray}
for any conformational function $A(\vec{q})$, we could linearly expand $\Omega_{X,eq}(\vec{q})$ with a complete set of conformational functions
(also referred to as basis functions in the following) $\{A^{\mu}(\vec{q})\}$, as
\begin{equation}
\label{eq:expand} \Omega_{X,eq}(\vec{q}) = 1+ \sum_{\mu,\nu} g_{\mu\nu}(P_{X}) \langle \delta_{X} A^{\mu}(\vec{q}) \rangle_{eq} \
\delta_{X}A^{\nu}(\vec{q}), 
\end{equation}
where, $\delta_{X}A^{\mu}(\vec{q}) \equiv A^{\mu}(\vec{q})-\langle A^{\mu}(\vec{q}) \rangle_{X}$ and $g_{\mu\nu}(P_{X})$ is the inverse of the
variance-covariance matrix, $g^{\mu\nu}(P_{X})$, which is defined as $\langle \delta_{X}A^{\mu}(\vec{q}) \delta_{X}A^{\nu}(\vec{q}) \rangle_{X}$
with $\langle \cdots \rangle_{X}$ denoting the average over $X$ samples. In Eq.~(\ref{eq:expand}), the equilibrium ensemble average $\langle
\delta_{X}A^{\mu}(\vec{q})\rangle_{eq}$ actually relies on the trajectory weights $\{w_i\}$ through Eq.~(\ref{eq:weight}). Thus, substituting
Eq.~(\ref{eq:expand}) into Eq.~(\ref{eq:realw}) gives out the following self-consistent linear equations of $\{w_i\}$, 
\begin{eqnarray}
w_{i} = 1 + \sum_{j} \Gamma_{ij} w_{j}, 
\label{eq:linear-equation}
\end{eqnarray}
where $\Gamma_{ij} = \frac{1}{p} \sum_{\mu,\nu} g_{\mu\nu}(P_{X}) \langle \delta_{X} A^{\mu}(\vec{q}) \rangle_{i^{+}} \langle \delta_{X}
A^{\nu}(\vec{q}) \rangle_{j}$, $i,j = 1, \cdots, p$. $p$ is the number of trajectories, and we already set $\sum_{i} w_{i} = p$. All the
parameters in Eq.~(\ref{eq:linear-equation}) could be calculated simply by averaging basis functions over simulation trajectories. Since the
current formulation is exempted from any information about $P_{init}(\vec{q})$, initial conformations could be arbitrarily selected from
different sources, such as coarse-grained simulation, high-temperature simulation, experimental knowledge, or even theoretical conjecture, then the conformational space could be more efficiently traversed and structured.

Equation~(\ref{eq:linear-equation}) could be written as $\vec{G}^{i} \cdot \vec{w} =0$, $(i=1, \ldots, p)$, where $\vec{G}^{i}=(G_{i1}, \ldots,
G_{ip})^{T}$ is the vector with components $G_{ij} \equiv \Gamma_{ij} - \delta_{ij} +\frac{1}{p}$, $(i,j=1, \ldots, p)$, and $\delta_{ij}$ is
the Kronecker delta symbol. $\vec{w}=(w_1,\ldots, w_p)^T$ is the vector of trajectory weights, which could be thought as the normal vector of
the hyper-plane spanned by the trajectory-mapped vectors $\{\vec{G}^{i}\}$. The equations could be further adjusted to the following form,
\begin{equation}
\label{eq:sym_lin_eq}
H \vec{w} =0,
\end{equation}
by minimizing the residual $I=\sum_{i} (\vec{G}^{i} \cdot \vec{w} - 0)^{2}$. Here $H=G^{T} G$ is the covariance matrix of the vectors
$\{\vec{G}^{i}\}$. Compared to Eq.~(\ref{eq:linear-equation}), Eq.~(\ref{eq:sym_lin_eq}) is easier for analytical treatment due to the symmetric
form of $H$. It is also more general in application. For example, when multiple trajectories are generated from one initial conformation for
better statistics, thus the number of trajectories is larger than the number of independent weights; Eq.~(\ref{eq:sym_lin_eq}) still works.

There are a few key points in the WED method. 
(1) $H$ is a positive semi-definite matrix with at least one zero eigenvalue. If the conformational space is well connected by simulation
trajectories, the ground state of $H$ will be non-degenerate, which uniquely determines the trajectory weights. Otherwise, if the simulation
trajectories are isolated in different conformational regions by large free energy barriers, $H$ will have degenerate ground state ({\it i.e.}
multiple zero eigenvalues), leaving the relative weights between trajectories in different regions indeterminable. When a few trajectories
connecting the different regions exist, the zero eigenvalues of $H$ will be perturbed to small nonzero values. The small-eigenvalue eigenvectors
of $H$ could be manipulated to extract the information of conformational states and transitions between them. This data mining process is
usually performed by projecting $\{\vec{G}^{i}\}$ to the small-eigenvalue eigenvectors. Since $\{\vec{G}^{i}\}$ one to one corresponds to the
simulation trajectories, the projection effectively maps the trajectories into low-dimensional space for subsequent classification. In the
following, we align the eigenvectors of $H$ by their eigenvalues with an ascending order, \emph{e.g.}, the first eigenvector is the one
corresponding to the smallest eigenvalue of $H$. The larger-eigenvalue eigenvectors mainly reflect the intrastate statistical fluctuation among
trajectories.

(2) Although the expansion in Eq.~(\ref{eq:expand}) is exact if and only if the set of conformational functions is complete, it is not necessary
to include too many basis functions in WED. This is because only the mean values of $\Omega_{X,eq}(\vec{q})$ over a large number of conformation
samples, instead of the values of $\Omega_{X,eq}(\vec{q})$ for different conformations, are required in the construction of the linear equations
of $\{w_i\}$ [Eq.~(\ref{eq:linear-equation}) and (\ref{eq:sym_lin_eq})]. In practice, only physically relevant and important quantities of
the simulation system need to be selected as basis functions to distinguish conformational meta-stable states. The selection does not demand
much foreknowledge. For biological macromolecules, the important inner coordinates, such as dihedral angles and pair distances, could be chosen
as basis functions to characterize the conformational motion. Various physical quantities, such as the potential energy of the system,
solute-solvent interactions, could also be included for the searching of related kinetic or thermodynamic phenomena. The variance-covariance
matrix $g^{\mu\nu}$ will ensure the consistent consideration of different classes of basis functions, and provide the measure of their relative
importance (after orthogonalizing the basis functions based on $g^{\mu\nu}$).

(3) There is only one free parameter in WED method, ${\tilde t}^{*}=\tilde{t}/\tau$, for collecting the $X$ samples. Large ${\tilde t}^{*}$ may
bring systematical error to the estimated equilibrium properties, and small ${\tilde t}^{*}$ will reduce the sample number in $X$, leading to a 
larger statistical uncertainty. $\tilde{t}^{*}=0.01 \sim 0.1$ is usually applied in the current work with satisfiable results obtained. We also
point out that, for the purpose of discovering the states in conformational space, the results are not very sensitive to ${\tilde t}^{*}$.

\section{Results and Discussions}
\label{sec:results}

\subsection{System with one-dimensional glassy potential}
\label{sec:oned}

\begin{figure}
  \centerline{ \includegraphics[width=.4\textwidth]{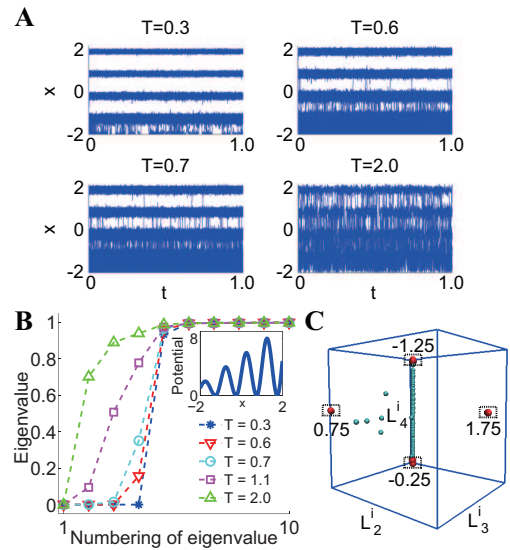} }
  \caption{(Color online) Trajectories, eigenvalues of $H$ and projection map of one-dimensional system. (A) Trajectories simulated at four different temperatures with a one-dimension glassy potential.
   All $400$ trajectories are plotted except for $T=2.0$, where only $20$ randomly selected ones are plotted. The horizontal axis denotes the
   reduced simulation time with the full simulation time scaled to $1.0$. The vertical axis denotes the coordinate of one-dimensional system.
   (B) The first ten eigenvalues of $H$ at five temperatures. The one-dimensional glassy potential is also shown as inset.
  (C) The projection of $\{\vec{G}^{i}\}$ to the second, third and fourth eigenvectors of $H$ ($T=0.6$).
   The points (red) representing non-transition trajectories in the four major potential wells $1.75$, $0.75$, $-0.25$ and $-1.25$ are
   gathering, respectively, inside the four regions enclosed by dashed-edge squares. The correspondence between states and regions is labeled in the graph.
  The other points (cyan) represent the transition trajectories between states. See the main text for more details.}
  \label{fig1}
\end{figure}

We first illustrate the WED method in a one-dimensional system with glassy potential. There are four major potential wells respectively located around
the positions $-1.25, -0.25, 0.75$, and $1.75$ in this system [see the inset of Fig.~\ref{fig1}(b)]. We name these potential wells with the
positions of their minima in the following. For each WED simulation, $400$ trajectories are generated with the initial conformations randomly
selected in the interval of $[-2.0,2.0]$. The system is investigated under five temperatures of $0.3$, $0.6$, $0.7$, $1.1$, and $2.0$. More
details are shown in the Appendix.

The simulation trajectories and the ten smallest eigenvalues of $H$ for WED analyses under different temperatures are shown in Fig.~\ref{fig1}.
At $T=0.3$, there is no simulation trajectory connecting the different potential wells. As temperature increases, transition events emerge, and
the total number of transition trajectories increases fast [see Fig.~\ref{fig1}(a)]. Correspondingly, four zero (or almost zero) eigenvalues of
$H$ are found at $T=0.3$ [see Fig.~\ref{fig1}(b)]. As temperature increases, the fourth smallest eigenvalue begins to deviate from zero due to the
transition trajectories between potential wells $-1.25$ and $-0.25$. At $T=0.7$, due to the several transition trajectories between potential
wells $-0.25$ and $0.75$, the third eigenvalue is also lifted to small positive value. Subsequently at $T=1.1$, except for the smallest one, all
the other eigenvalues are prominently deviating from zero. However, since only limited fraction of trajectories can pass the highest energy
barrier in the system, the second smallest eigenvalue is still relatively small (about $0.1$). Finally at $T=2.0$, there is only one near-zero
eigenvalue left, indicating the ergodicity of the simulation data as a whole.

\begin{figure}
  \centerline{ \includegraphics[width=.4\textwidth] {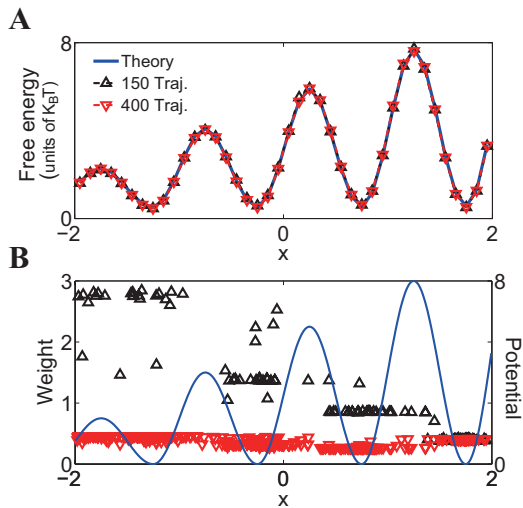} }
  \caption{(Color online) Reconstructed energy curve and the trajectory weights
  of one-dimensional system. (A) The energy curves reconstructed with $150$-trajectory
  (black upward triangles) and $400$-trajectory (red downward triangles) data set are
  shown together with the theoretical curve (blue solid line). The three lines
  almost overlap with each other.
  (B) Trajectory weights for $150$-trajectory (black upward triangles) and $400$-trajectory data set (red downward triangles) are shown.
  The horizontal axis denotes the initial position of trajectories.
  The weights has been scaled to ensure the equal average weight of
  trajectories from $1.75$ potential well for the two data sets. The potential energy is also
  shown as solid line.}
   \label{fig2}
\end{figure}

For WED simulation at $T=1.1$, since there is only one zero eigenvalue of $H$, we could uniquely determine the trajectory weights to reproduce
the equilibrium distribution (or energy curve) of the system. Although, only part of the WED trajectories are found to be able to pass the
highest potential barrier (either once or more), thus, single trajectory is far from ergodic, the energy curve of this system is correctly
reproduced by re-weighting the simulation trajectories [see Fig.~\ref{fig2}(a)]. For further testing, we build a $150$-trajectory subset from all
the $400$ trajectories, so that the distribution of the initial conformations of the $150$ trajectories is very different from that of all the
$400$ trajectories. In practice, the subset involves all the $75$ trajectories initially inside the $1.75$ potential well and $75$ ones that are
randomly chosen from the remaining $325$ trajectories initially outside the $1.75$ potential well. With the subset of trajectories, we
reconstruct Eq.~(\ref{eq:sym_lin_eq}), recalculate the trajectory weights, and use the weights and the $150$ trajectories to reproduce the
equilibrium distribution of the system. The energy curve is again correctly predicted as shown in Fig.~\ref{fig2}(a). More intuition could be
perceived from Fig.~\ref{fig2}(b), where the trajectory weights for both the original $400$-trajectory data set and the $150$-trajectory subset
are shown. For the $150$-trajectory data set, the fraction of conformations outside the $1.75$ potential well is reduced a lot, thus, the
trajectories which started outside the $1.75$ potential well are specified larger weights in response. The changing of trajectory weights offsets the
changing of the initial distribution of the trajectories, leading to the same reconstructed energy curve. Besides, the weight of a trajectory is
found to be mainly dependent on the potential well from where the trajectory is started, rather than the initial conformation of the trajectory,
consistent with our supposition in the derivation of Eq.~(\ref{eq:sym_lin_eq}).

At lower temperature of $T=0.6$, the ground state of $H$ is degenerate. We project $\vec{G}^{i}$ to the second, third and fourth eigenvectors of
$H$, which maps each trajectory $i$ to the point $(L^{i}_{2},L^{i}_{3},L^{i}_{4})$ in a three-dimensional space. Here $L^{i}_{\alpha} \equiv
\vec{G}^{i} \cdot \vec{u}_{\alpha}$, $\vec{u}_{\alpha}$ is the $\alpha$th eigenvector of $H$, and $L^{i}_{1}$ is always zero. In the
three-dimensional space, the $400$ trajectories could be classified as shown in Fig.~\ref{fig1}(c). There are four highly concentrated groups,
respectively, with $77$, $92$, $37$ and $66$ points of almost the same coordinates. The points in these groups are verified to represent the
non-transition trajectories in the potential wells $1.75$, $0.75$, $-0.25$, and $-1.25$, respectively. The $119$ points located along the line connecting $-1.25$ and $-0.25$ correspond to transition trajectories between potential wells $-1.25$ and $-0.25$, the other $9$ points located in the plane of $-1.25$, $-0.25$, and $0.75$ correspond to that among the wells $-1.25$, $-0.25$, and $0.75$.

\begin{figure}
  \centerline{\includegraphics[width=.4\textwidth]{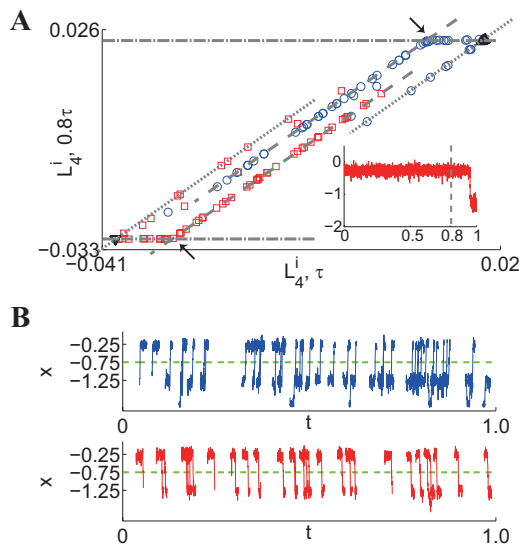} }
   \caption{(Color online) Trajectory identification and shortening for locating transition state ensemble in one-dimensional system. (A) $\{L^i_4\}$ calculated with truncated trajectories (at $0.8\tau$)
  versus those with the full trajectories. Only data for trajectories which started from the potential wells $-1.25$ and $-0.25$ are shown.
  The non-transition trajectories in $-1.25$ and $-0.25$ are shown as upward and downward triangles respectively,
  and the transition trajectories which started from the two states are plotted as (blue) circles and (red) squares, respectively.
  The straight lines are plotted by hand, and the arrows indicate the intersection points between dash-dotted horizontal lines and dashed-inclined lines.
  The inset illustrate the truncation of one simulation trajectory at the normalized time $t/\tau=0.8$.
    (B) For the supposed single-transition trajectories, short ($0.02 \tau$) trajectory segments centered by the predicted transition times are shown.
  The upper panel shows the predicted transitions from the potential wells $-1.25$ to $-0.25$; 
  the lower panel shows the transitions predicted to be in the reversed direction. The horizontal axes denote the reduced simulation time with the full simulation
  time scaled to $1.0$.}
  \label{fig3}
\end{figure}

With the classification of trajectories, the equilibrium properties in each potential well could be estimated by the non-transition trajectories
inside. Moreover, considering the $119$ transition trajectories between potential wells $-1.25$ and $-0.25$, it is possible to further extract
more detailed information of the super-state containing these two states. We truncate the ending segment of each trajectory and repeat the
analysis ({\it i.e.} reconstructing $H$, analyzing its spectral properties etc.). For the trajectories in the super-state containing $-1.25$ and
$-0.25$, we plot the $\{L^{i}_{4}\}$ calculated with truncated trajectories (at $0.8 \tau$) versus those with the full trajectories in
Fig.~\ref{fig3}(a). In this figure, a one-to-one correspondence between the data points and the simulation trajectories exists, which helps to
further classify the trajectories as follows. (i) The data points could be approximately separated into two groups. Those in the same group
correspond to the trajectories which started from the same state (either $-1.25$ or $-0.25$). (ii) The data points with extremal $L^i_4$ values along
both axes correspond to the non-transition trajectories. No matter whether truncated or not, these trajectories are always non-transition ones, and
should always be similar to each other in $L^i_4$ value. (iii) Except for the points corresponding to the non-transition trajectories, the other
points in the dash-dotted horizontal lines in Fig.~\ref{fig3}(a) correspond to the transition trajectories where transition does not happen before
$0.8\tau$. These trajectories become non-transition ones only after truncation and should have similar truncation-calculated $L^i_4$ values. 
Besides, we also find that, (iv) the data points in the dashed-inclined lines correspond to trajectories with an odd number of transitions which
all happen within $0.8 \tau$. (v) The ones in the dotted-inclined lines correspond to trajectories with even number of transitions which all
happen within $0.8 \tau$. (vi) All the other points outside the straight lines in Fig.~\ref{fig3}(a) correspond to the trajectories with early
transitions occurred within $\tilde{t}$, and the multiple transition trajectories with transition happened both within $0.8 \tau$ and after $0.8
\tau$. Noticing that the truncation of trajectories actually adjusts the occupation fraction of the trajectories in different states, the
observed allocation from (iv) to (vi) could be explained by supposing the existence of the linear relation between $\{L^i_4\}$ and the
occupation fraction of trajectories in different states.

Since there are considerable fraction of non-transition trajectories within $\tau$, we could safely assume that
almost all the odd-transition trajectories are actually single-transition trajectories. Thus, their transition time should be linearly dependent
on their occupation fraction in different states. Therefore, the linear relation between transition time and $L^i_4$ value should exist for
these putative single-transition trajectories. To predict the linear relation, there are two key points in Fig.~\ref{fig3}(a), corresponding to
the two intersection points between dash-dotted horizontal lines and dashed-inclined lines. These two points correspond to single-transition
trajectories, which started from different potential wells ($-1.25$ and $-0.25$), and happen transition at $0.8\tau$. By truncating the
trajectories to different lengths ({\it e.g.}, $0.5 \tau$, $0.6 \tau$, $0.7 \tau$, $0.8 \tau$ and $0.9 \tau$), and reproducing the analogous figures to Fig.~\ref{fig3}a, a few pairs of key points could be collected. Two
sets of linear relations between transition time and $L^i_4$ value for the supposed single-transition trajectories could be determined, which
apply, respectively, to the $-1.25$-started trajectories (trajectories started from the $-1.25$ potential well) and the $-0.25$-started
trajectories.

For all the transition trajectories in the dashed-inclined lines and dash-dotted horizontal lines in Fig.~\ref{fig3}(a), $0.02 \tau$ trajectory
segments centered by the predicted transition times are shown in Fig.~\ref{fig3}(b). Except for one trajectory, which is non-transition in
$0.8\tau$ and happens twice transitions after $0.8\tau$, the other trajectories all pass the potential barrier between $-1.25$ and $-0.25$
within the predicted time windows. We could expect that for systems with well-defined transition states, it is always
possible to efficiently shorten the transition trajectories to locate the transition state ensemble. Similar example for the
solvated alanine dipeptide system will be shown in Sec.~\ref{sec:alanine}.

\subsection{System of solvated alanine dipeptide}
\label{sec:alanine}

\begin{figure}
  \centerline{\includegraphics[width=.4\textwidth]{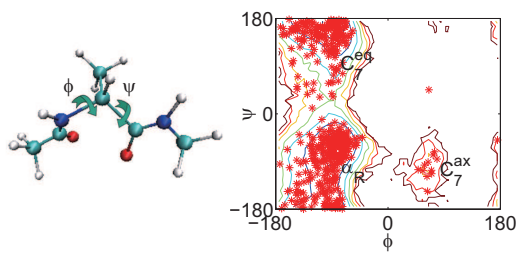} }
   \caption{(Color online) Illustration of alanine dipeptide molecule and the selected initial conformations from $600 {\rm K}$
   simulation. In the left panel, the alanine dipeptide molecule is shown with the two
   important dihedral angles, $\phi$ and $\psi$, labeled. In the right panel, the $500$ initial conformations of WED
   simulation are projected onto the $\phi$-$\psi$ plane (red stars), taking their dihedral angles as coordinates.
   The background is the free energy surface on $\phi$-$\psi$ plane reconstructed with the WED analysis of $450 {\rm K}$ data.}
  \label{fig4}
\end{figure}

The alanine dipeptide molecule is shown in Fig.~\ref{fig4} (left panel). There are only two important main chain dihedral angles  $\phi$ and
$\psi$  in this system. The 22-atom molecule is solvated in $522$ TIP3P waters. $500$ conformations are first collected from a $10 {\rm ns}$
simulation of the system at $T= 600 {\rm K}$. The projections of these conformations to the $\phi-\psi$ plane are shown in Fig.~\ref{fig4} (right
panel). These conformations are mainly located in the three free-energy wells on the $\phi-\psi$ plane, \emph{i.e.},  $C^{eq}_{7}$ and $\alpha_{R}$ with
negative $\phi$ value ($488$ in $500$) and $C^{ax}_{7}$ with a positive $\phi$ value ($12$ in $500$). $C^{ax}_{7}$ is less stable compared to
$C^{eq}_{7}$ and $\alpha_{R}$. Starting from each of these conformations, the system is simulated for $300 {\rm ps}$ (or $600 {\rm ps}$). More
details are shown in the Appendix.

\begin{figure}
   \centerline{\includegraphics[width=.4\textwidth]{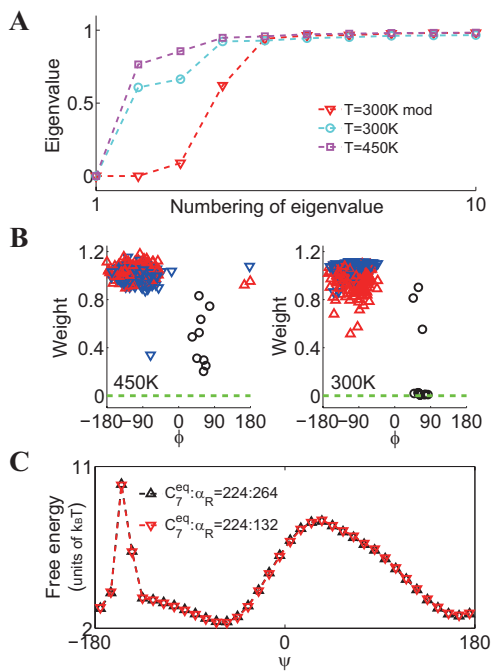}}
  \caption{(Color online) Eigenvalues of $H$, trajectory weights, and reconstructed free-energy surface for
  solvated alanine dipeptide system. (A) The eigenvalues of $H$ for solvated alanine dipeptide system with different temperatures and force fields.
  `$300 {\rm K}$ mod'
  labels the results of $300 {\rm K}$ simulation with modified potential.
   (B) The calculated trajectory weights for simulations at $450 {\rm K}$ (left panel) and $300 {\rm K}$ (right panel) with standard force field. The horizontal
  axis denotes the $\phi$ angle of the initial conformations of trajectories.
  The trajectories started from $C^{eq}_7$, $\alpha_R$ and $C^{ax}_7$ regions are plotted as (red) upward triangles, (blue) downward triangles and (black) circles, respectively.
  (C) The free energy curve along $\psi$ axis constructed by the weighted trajectories simulated with modified potential at $300 {\rm K}$.
  The curve labeled with (black) upward triangles is obtained with all the $488$ trajectories in $C^{eq}_7$($224$ in $488$) and $\alpha_R$($264$ in $488$); the one
  labeled with (red) downward triangles is constructed with half of the trajectories started from $\alpha_R$ omitted.}
 \label{fig5}
\end{figure}

There exists only one zero eigenvalue of $H$ under both temperatures of $T=450 {\rm K}$ and $T=300 {\rm K}$ [see Fig.~\ref{fig5}a], suggesting
that the equilibrium properties of the system could be reproduced with current simulation data. The trajectory weights estimated by WED are
shown in Fig.~\ref{fig5}(b). At $T=450 {\rm K}$, the weights of the $C_{7}^{eq}$-started trajectories (trajectories started from $C_{7}^{eq}$)
have similar mean value and fluctuation with those of the $\alpha_{R}$-started trajectories. In contrast, the weights of the
$C_{7}^{ax}$-started trajectories are partially depressed to smaller values.

At $T=300 {\rm K}$, the weights of the $C_{7}^{eq}$-started trajectories, as a whole, are slightly smaller than those of the
$\alpha_{R}$-started trajectories [see Fig.~\ref{fig5}b, right panel]. With current simulation data, the life time of $C_{7}^{eq}$ and
$\alpha_R$ are estimated to be $30.5$ and $25.2$ {\rm ps} at $T=300 {\rm K}$ (the corresponding values are $9.56$ and $10.3
{\rm ps}$ at $T=450 {\rm K}$). Consequently, some of the $300{\rm ps}$ trajectories may not be able to reach equilibrium between these two
states. Thus, to reconstruct the equilibrium distribution in the region containing $C_{7}^{eq}$ and $\alpha_R$, it is inevitable to specify
diversified weights to different trajectories, which explains the slight difference in trajectory weights between the $C_{7}^{eq}$-started and
the $\alpha_R$-started trajectories. In comparison with $T=450 {\rm K}$, the weights of the $C_{7}^{ax}$-started trajectories are further
depressed in response to the further instability of $C_{7}^{ax}$ at lower temperature.

\begin{figure}
  \centerline{\includegraphics[width=.4\textwidth]{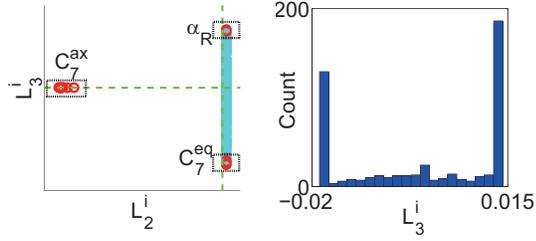} }
  \caption{(Color online) Projection map of solvated alanine dipeptide system simulated at $300 {\rm K}$ with modified force field. In the left panel,
  the projection of $\{\vec{G}^{i}\}$ to the second and third eigenvectors of $H$ are shown.
  The points (red circles) representing the non-transition trajectories in the three free-energy wells of $C_{7}^{eq}$, $\alpha_R$ and $C_{7}^{ax}$ are
  gathering, respectively, inside the three regions enclosed by dashed-edge squares. The correspondence between states and regions is labeled in the graph.
  The other points (cyan circles) represent the transition trajectories between states. See the main text for more details. In the right panel, the histogram
  of $L^i_3$ value for trajectories in $C_{7}^{eq}$ and $\alpha_R$ is shown to illustrate the concentration of trajectories
  at the extremal values of $L^i_3$.}
  \label{fig6}
\end{figure}

Adding potential barriers onto the standard dihedral energy terms of $\phi$ and $\psi$, we re-simulate $500$ trajectories with $\tau=600 {\rm
ps}$ and analyze these trajectories with WED method. These potential barriers will kinetically further separate the three free-energy wells of
$C_{7}^{eq}$, $\alpha_{R}$, and $C_{7}^{ax}$ (see the Appendix for more details). As a result, two zero (or near zero) eigenvalues of $H$ are
found [see Fig.~\ref{fig5}(a)], predicting two groups of simulation trajectories secluded in different free-energy wells. Consistently, the
projection values of $\{\vec{G}^i\}$ to the second eigenvector of $H$ ({\emph i.e.}, $\{L^i_2\}$) classify all the 500 trajectories into two groups
(see Fig.~\ref{fig6}, left panel). While $12$ trajectories in one group are  located inside $C^{ax}_{7}$, the remaining $488$ trajectories could be
further classified based on their $L^{i}_{3}$ values, considering the small value of the third eigenvalue of $H$. These $488$ trajectories are
identified as non-transition trajectories in the two states, $C^{eq}_{7}$ and $\alpha_{R}$, and transition trajectories between them. In the
$L^i_2$-$L^i_3$ plane, they lie along a straight line almost parallel to the $L^i_3$ axis, with the non-transition trajectories gathering at the
two terminals of the line (with extremal $L^i_3$ values, see Fig.~\ref{fig6}, right panel). By independently analyzing the $488$ trajectories,
the weights of these trajectories could be obtained to construct the equilibrium distribution of the super-state containing $C^{eq}_{7}$ and
$\alpha_{R}$. The resulting free energy profile along $\psi$ is shown in Fig.~\ref{fig5}(c). We randomly throw away half of the trajectories that 
started from $\alpha_{R}$ ({\it e.g.}, $132$ of $264$ trajectories), and redo the WED analysis for the remaining $356$ trajectories. Although the
conformations in $\alpha_{R}$ have been (approximately) reduced by half, the estimated weights of the remaining $\alpha_{R}$-started
trajectories are indeed almost doubled relative to the previous values. Consequently, the reconstructed free energy profile closely matches the
previous one, suggesting the robustness of WED method.

\begin{figure}
  \centerline{ \includegraphics[width=.4\textwidth]{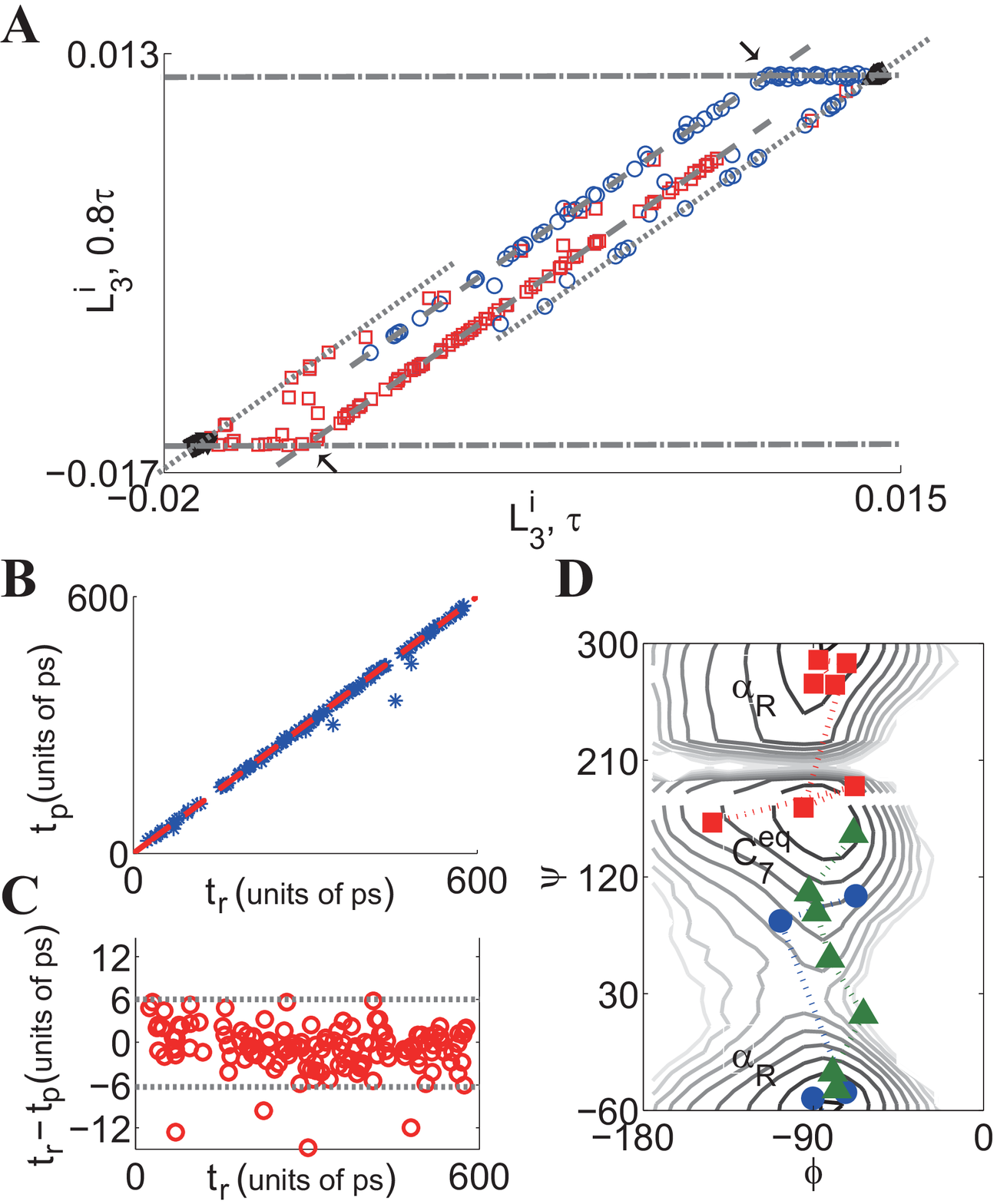} }
  \caption{(Color online) Trajectory identification and shortening for locating transition states in the system of solvated alanine dipeptide.
  (A) $L^i_3$ calculated with truncated trajectories (at $0.8{\rm \tau}$) versus those with full trajectories for the simulations
  with modified potential at $T=300 {\rm K}$. Only data for trajectories which started from $\alpha_R$ and $C^{eq}_7$ are shown.
  The non-transition trajectories in $\alpha_R$ and $C^{eq}_7$ are shown as upward and downward triangles, respectively,
  and the transition trajectories started from the two states are plotted as (blue) circles and (red) squares, respectively.
  The straight lines are plotted by hand,
  and the arrows indicate the intersections between dash-dotted horizontal lines and dashed inclined lines.
   (B) The predicted transition time, $t_{p}$,
   for the supposed single-transition trajectories versus their real transition time, $t_{r}$.
   The $t_{p}=t_{r}$ relation is plotted as (red) dashed line for comparison.
  (C) The error of the predicted transition time versus the real transition time.
    (D) Illustration of transition paths.
    Three $6 {\rm ps}$-length segments of trajectories around their predicted transition time are shown with different symbols.
   The background is the free energy surface reconstructed from $300 {\rm K}$ simulation trajectories with modified potential.}
  \label{fig7}
\end{figure}

Similar to the one-dimensional system with glassy potential, for the $488$ trajectories in $C^{eq}_{7}$ and $\alpha_{R}$ potential wells, we also find
the linear relation between the $L^i_3$ value and the occupation fraction of trajectories. Analogous figure to Fig.~\ref{fig3}(a) is plotted as
Fig.~\ref{fig7}(a). The predicted transition times for the putative single-transition trajectories [trajectories with their representing points
located on the dashed inclined lines in Fig.~\ref{fig7}(a)] are in great agreement with their real transition times directly identified along the
simulation trajectories [see Figs.~\ref{fig7}(b) and \ref{fig7}(c)]. The fraction of the trajectories which happen transition within the $6$ {\rm ps} ($12$ {\rm ps}) time windows centered by their predicted transition times, reaches $75\%$($93\%$) percent. For illustration, the predicted $6 {\rm ps}$
segments of three trajectories are shown in Fig.~\ref{fig7}(d). 

\section{Conclusion}
\label{sec:conclude} We present the WED scheme for systematically exploring the kinetic state structure of conformational space. WED works by
automatically discovering the mutual relation between parallel generated trajectories. It could also combine partially overlapped trajectories
in conformational space to estimate the equilibrium properties of complex systems with rugged potential-energy surface. The method works well
without existing experimental or simulation knowledge of the system, and may not suffer much from increasing system dimension by only applying
relevant physical quantities as basis functions. Exempted from the knowledge of the initial conformational distribution, WED provides
flexibility to choose as dispersive as possible start points of trajectories in the whole conformational space. For example, initial
conformations could come from simulations at different temperatures, as well as theoretically and experimentally important conformations ({\it
e.g.}, completely or partially folded structures of proteins). After arbitrarily adding back the missing degrees of freedom and short relaxation,
it is also possible to adopt coarse-grained simulation conformations as starting points of WED simulation. Since the detailed sub-states in
shorter time scales could be detected by sequentially chopping the trajectories and repeating our analysis, the hierarchical state structure of the 
free-energy landscape~\cite{KrivovK2004} could be distilled out up to the total simulation time scale. Finally, for slow transition dynamics,
which is impractical to be realized by simply increasing trajectory number, the combination of WED with current techniques for studying
two-point slow dynamics~\cite{BolhuisCDG2002,Elber2002,ERV2005} should be interesting.

{\bf Acknowledgements} 
Authors acknowledge the Max Planck Society(MPG) and the Korea Ministry of Education, Science and Technology(MEST) for the
support of the Independent Junior Research Group at the Asia Pacific Center for Theoretical Physics (APCTP). X.Z. is grateful to Y. Jiang for stimulating discussions.

\section{Appendixes}
\label{sec:method}

\subsection{Simulation and analysis method for one-dimension system}

For one-dimension system with coordinate $x$ and potential function $U(x)$, the over-dampened Langevin equation
\begin{equation}
\frac{dx}{dt}=-\frac{1}{\gamma}\frac{dU}{dx}+\sqrt{\frac{2k_BT}{\gamma}}\xi(t).
\end{equation}
is adopted to generate the dynamical trajectories of $x$. Where $\gamma$ is the frictional coefficient, $k_B$ is the Boltzmann constant, $T$ is
the simulation temperature, and $\xi(t)$ is the white noise satisfying $\langle\xi(t)\xi(t')\rangle=\delta(t-t')$ with $\langle\rangle$ denoting
ensemble average of noise. We simply take $k_B$ and $\gamma$ as unity to get the dimension-reduced units for time, position and temperature.
Reflecting boundary condition is assigned for all the simulations.

The analytical expression of one-dimension glassy potential is as following
\begin{displaymath}
U(x)=\left\{ \begin{array}{lrcl} \infty & &x&<-2.0\\
1+\sin(2\pi{}x) & -2.0<&x&<-1.25\\
2[1+\sin(2\pi{}x)] & -1.25<&x&<-0.25\\
3[1+\sin(2\pi{}x)] & -0.25<&x&<0.75\\
4[1+\sin(2\pi{}x)] & 0.75<&x&<1.75\\
5[1+\sin(2\pi{}x)] & 1.75<&x&<2.0\\
\infty & &x&>2.0
\end{array} \right..
\end{displaymath}
For ED simulations, $400$ initial positions are randomly selected in $[-2.0,2.0]$ interval. Each trajectory is simulated for $50$ time length,
and the frames are recorded every $0.01$ time length. In analysis, $t^{*}=0.02$ is chosen, corresponding to $40000$ conformations in the $X$
sample.

The one-dimension trigonometrical functions
\begin{equation}
\label{eq:1dp_basis} \cos(\frac{\pi{}mx}{2}),\sin(\frac{\pi{}nx}{2})\quad{}m,n=1,2,3\ldots.
\end{equation}
are selected as basis functions in analysis. The first $20$ basis functions, e.g. $m$, $n$ both from $1$ to $10$, are included in calculation.
Similar results could be obtained with $10$ basis functions, with $m$ and $n$ both from $1$ to $5$. To reconstruct the energy curve by WED
method, a $40$-bin histogram is generated. The theoretical curve is calculated by integrating the theoretical equilibrium distribution inside
each bin.


\subsection{Simulation and analysis method for alanine dipeptide with explicit solvent}

The alanine dipeptide, a $22$-atom small peptide molecule, is solvated in $522$ TIP3P water molecules with totally $1588$ atoms in the system.
All the simulations are performed under $NVT$ ensemble (constant particle number, constant volume and constant temperature) with NAMD simulation
package\cite{NAMD2005} and Charmm27 force field. The Langevin thermostat is chosen to keep the temperature of the system, with a damping
coefficient of $5{\rm ps}^{-1}$. Periodic boundary condition is imposed. The system has a box size of $25.9{\rm \AA}\times{}24.5{\rm
\AA}\times{}27.8{\rm \AA}$, and PME method is applied to calculate the electrostatic energy with Particle Mesh Ewald PME grid size chosen as $32$ along all the
directions. Van de Waals interaction is cut off at $12{\rm \AA}$ and switched to zero from $10{\rm \AA}$. In the WED simulations, the first
$1{\rm ps}$ simulation for each trajectory is taken as relaxation, then each trajectory is simulated for $300 {\rm ps}$ with conformations
recorded every $0.5{\rm ps}$ ($300{\rm K}$ and $450${\rm K} simulation with standard force field) or $600 {\rm ps}$ with conformations recorded
every $0.2{\rm ps}$ ($300{\rm K}$ simulation with modified force field).

To mimic a disconnected free energy landscape under $300 {\rm K}$ temperature for this system, we added the following style of boundary potential
on the Charmm27 force field for selected dihedral angles.
\begin{displaymath}
f(\chi,\chi_{low},\chi_{high},\Delta\chi,E_0)= \left\{
\begin{array}{lrcl}
\frac{\cos(\chi-\chi_{low})-\cos(\Delta\chi)}{1-\cos(\Delta\chi)}E_0 & \chi_{low}-\Delta\chi<&\chi&<\chi_{low}\\
E_0 & \chi_{low}<&\chi&<\chi_{high}\\
\frac{\cos(\chi-\chi_{high})-\cos(\Delta\chi)}{1-\cos(\Delta\chi)}E_0 & \chi_{high}<&\chi&<\chi_{high}-\Delta\chi\\
0 & \chi_{high}<&\chi&<\chi_{low}+360
\end{array} \right.
\end{displaymath}
Where, $\chi$ is the degree of freedom on which the modified potential acts. $\chi_{low}$ and $\chi_{high}$ are the lower and upper boundary for
the potential function. Within the boundary, the potential function is identically equal to $E_0$. Out of the boundary, the potential function
will reduce to zero within $\Delta\chi$ length. The angles are measured with degree, and the energy unit is ${\rm Kcal/mol}$. Three potential
functions, $f(\psi,-155,-150,10,8.5)$, $f(\phi,0,10,15,12)$ and $f(\phi,130,140,15,4)$ are added by NAMD simulation package to modify the free
energy surface. $f(\psi,-155,-150,10,8.5)$ adjusts one of the two free energy barriers between $C^{eq}_7$ and $\alpha_R$, lifting the one at
$-150$ degree of $\psi$ angle to approximately the same height as the one at $25$ degree. The other two free energy functions are used to
separate $C^{ax}_7$ with the other free energy wells in conformational space.

In the WED analysis, we first choose physically important quantities as basis functions. The selected physical quantities include the potential
energy of the whole system, the square of potential energy, the sum of all dihedral energies in alanine dipeptide, the sum of all bonded
energies (bond, angle and dihedral energies) in alanine dipeptide. The physically selected basis functions are then complemented by two
dimensional trigonometrical functions of dihedral angles $\phi$ and $\psi$. These functions are
\begin{eqnarray}
\label{eq:hd_base_1} &\sin{}[(m+n)\phi],\cos[(m+n)\phi],m+n>0&\nonumber\\
&\sin{}(m\phi{})\sin{}(n\psi),\sin{}(m\phi{})\cos{}(n\psi),\cos{}(m\phi{})\sin{}(n\psi),\cos{}(m\phi{})\cos{}(n\psi),m\geq{}1,n\geq{}1&
\end{eqnarray}
We define the summation of $m$ and $n$ in equation (\ref{eq:hd_base_1}) as the order of these functions, and adopt the one to four order
functions in our analysis. Together with the physical quantities, there are $44$ basis functions included in calculation. However, the results
obtained with or without the physical basis functions are basically the same. The $12$ basis functions of one and two order trigonometrical
functions could already reproduce the results in current paper. The $t^*$ value of $0.05$ is adopted to analyze the simulation data at $450{\rm
K}$ and $300{\rm K}$ with standard force field. The one of $0.02$ is adopted to analyze the simulation data at $300{\rm K}$ with modified force
field. To reconstruct the free energy surface, the interval of $[-180, 180]$, either for $\phi$ or $\psi$, is divided into $36$ bins for
histogram construction.

\newpage

\newpage

\begin{thebibliography} {99}

\bibitem{BergN1991} B.~A. Berg and T.~Neuhaus, Phys. Lett. B {\bf 267}, 249 (1991).

\bibitem{WangL2001b} F.~Wang and D.~P. Landau, Phys. Rev. Lett. {\bf 86}, 2050 (2001).


\bibitem{SugitaO2000} Y.~Sugita and Y.~Okamoto, Chem. Phys. Lett. {\bf 329}, 261 (2000).

\bibitem{Voter1997b} A.~F. Voter, Phys. Rev. Lett. {\bf 78}, 3908 (1997).

\bibitem{ZhouJKZR2006} X.~Zhou, Y.~Jiang, K.~Kremer, H.~Ziock, and S.~Rasmussen, Phys. Rev. E {\bf 74}, 035701(R) (2006).

\bibitem{BolhuisCDG2002} P.~G. Bolhuis, D.~Chandler, C.~Dellago, and P.~L. Geissler, Annu. Rev. Phys. Chem. {\bf 53}, 291 (2002).

\bibitem{Elber2002} R.~Elber, A. Ghosh, and A. Cardenas, Acc. Chem. Res. {\bf 35}, 396 (2002).

\bibitem{ERV2005} Weinan E, Weiqing Ren, and E. Vanden-Eijnden, J. Phys. Chem. B {\bf 109}, 6688 (2005).

\bibitem{ThirumalaiMK1989} D.~Thirumalai, R.~D. Mountain, and T.~R. Kirkpatrick, Phys. Rev. A {\bf 39}, 3563 (1989).

\bibitem{Voter1998} A.~F. Voter,  Phys. Rev. B {\bf 57}, 13985(R) (1998).

\bibitem{ShirtsP2001} M.~R. Shirts and V.~S. Pande, Phys. Rev. Lett. {\bf 86}, 4983 (2001).

\bibitem{HuangBGP2008} X.~Huang, G.~R. Bowman, and V.~S. Pande, J. Chem. Phys. {\bf 128}, 205106 (2008).



\bibitem{StillingerW1982}  F.~H. Stillinger and T.~A. Weber, Phys. Rev. A {\bf 25}, 978 (1982).


\bibitem{BeckerK1997} O.~M.~Becker and M.~Karplus, J. Chem. Phys. {\bf 106}, 1495 (1997).

\bibitem{Wales2001} D.~J. Wales, Science {\bf 293}, 2067 (2001).


\bibitem{KrivovK2004} S.~V. Krivov and M.~Karplus, Proc. Natl. Acad. Sci. USA {\bf 101}, 14766 (2004).


\bibitem{RaoC2004} F.~Rao and  A.~Caflisch, J. Mol. Biol. {\bf 342}, 299 (2004).

\bibitem{GfellerRCR2007} D.~Gfeller, P.~D.~L. Rios, A.~ Caflish, and F.~Rao, Proc. Natl. Acad. Sci. U.S.A. {\bf 104}, 1817 (2007).

\bibitem{ChoderaSPSD2007} J.~D. Chodera, W.~C. Swope, J.~W. Pitera, C.~Seok, and K.~A. Dill, J. Chem. Theory  Comput. {\bf 3}, 26 (2007).




\bibitem{NAMD2005}
J.~C. Phillips, et al., J. Comput. Chem. {\bf 26}, 1781 (2005).

\end{thebibliography}
\end{document}